\documentclass[
reprint,twocolumn,
 amsmath,amssymb,
 aps,
]{revtex4-2}

\usepackage{graphicx}
\usepackage{dcolumn}
\usepackage{bm}

\usepackage{caption}
\usepackage{subcaption}

\usepackage{booktabs}
\usepackage{float}
\usepackage{indentfirst,csquotes}
\begin{document}


\title{Constraints for rare electron-capture decays mimicking detection of dark-matter particles in nuclear transitions}

\author{Aagrah Agnihotri} 
\email{aagrah.a.agnihotri@student.jyu.fi}
\affiliation{University of Jyv\"askyl\"a, Department of Physics, P.O. Box 35, FI-40014 Jyv\"askyl\"a, Finland}%
\author{Jouni Suhonen}
\email{jouni.t.suhonen@jyu.fi}
\altaffiliation[Also at]{\ International Centre for Advanced Training and Research in Physics (CIFRA), P.O. Box MG12, 077125 Bucharest-M{\u a}gurele, Romania}
\affiliation{University of Jyv\"askyl\"a, Department of Physics, P.O. Box 35, FI-40014 Jyv\"askyl\"a, Finland}

\author{Hong Joo Kim}
\email{hongjooknu@gmail.com}
\affiliation{Department of Physics, Kyungpook National University, Daegu 41566, Republic of Korea
}%
\date{\today}

\begin{abstract}
We give for the first time, theoretical estimates of unknown rare electron-capture (EC) decay branchings of $^{44}$Ti, $^{57}$Co, and $^{139}$Ce, relevant for searches of (exotic) dark-matter particles. The nuclear-structure calculations have been done exploiting the nuclear shell model (NSM) with well-established Hamiltonians and an advanced theory of $\beta$ decay. In the absence of experimental measurements of these rare branches, these estimates are of utmost importance for terrestrial searches of dark-matter particles, such as axionic dark matter in the form of axion-like particles (ALPs), anapole dark matter, and dark photons in nuclear transitions. Predictions are made for EC-decay rates of 2$^{nd}$-forbidden unique (FU) and 2$^{nd}$-forbidden non-unique (FNU) EC transitions that can potentially mimic dark-matter-particle detection in dedicated underground experiments designed to observe the absence of the corresponding nuclear electromagnetic transitions.
\end{abstract}
\maketitle

The study of nuclear $\beta$ decays has important implications for Beyond the Standard Model Physics (BSMP) \cite{Neutrinonuclearres,NBDforBSMP,NPSinNNBD,Guadilla_2024,Pagnanini}. Among different avenues of searches for BSMP, where the study of nuclear $\beta$ decays plays an important role, experimental searches for solar axions have long been among such endeavors \cite{axionmain,Axionreview}. Axions being potential dark-matter candidates \cite{PRESKILL1983127,ABBOTT1983133,DINE1983137,axiondm}, were originally hypothesized to exist as a solution to the strong CP problem \cite{CP1,CP2,CP3} and later generalized to classes of axion-like particles (ALPs), see the review \cite{Choi2021}. 

Cosmological and astrophysical constraints for ALPs are stringent, but it is also necessary to perform laboratory experiments in searches for axions or ALPs since these experiments are well-controlled, reproducible, and less model-dependent. There are several approaches of laboratory experiments based on axion-photon coupling, such as helioscopes, light-shining through walls interferometry, and solar-axion experiments as shown in the review \cite{axiondm}. Solar-axion experiments have been devised to detect the associated solar axions resonantly in terrestrial laboratories, like in the case of $^7$Li \cite{Belli2012}, $^{57}$Fe \cite{Derbin2011}, and $^{83}$Kr \cite{Jakovcic2004}.
To cover higher mass regions with terrestrial sources, collider, and beam-dump facilities \cite{Bauer2019}, nuclear reactor facilities \cite{Dent2020} and radio-active-source based experiments are ongoing or proposed.

\begin{figure}
\centering
\caption{Decay schemes for the three sources considered in this Letter. M1 transitions are marked by bold red arrows for all decay schemes. E1 and M2 transitions in (b) are marked by bold dark green and orange arrows, respectively. The E2 transition in (c) is marked by bold brown arrow.}
\label{fig:three_graphs}
     \begin{subfigure}[b]{0.5\textwidth}
         \centering
         \caption{EC decay scheme of $^{139}$Ce.}
         \includegraphics[width=\textwidth]{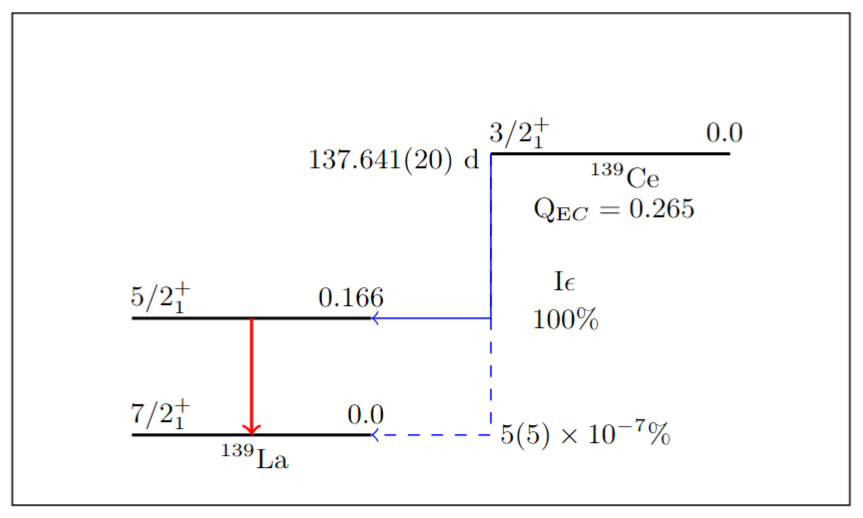}
         \label{fig:139ce}
     \end{subfigure}
     \begin{subfigure}[b]{0.5\textwidth}
         \centering
         \caption{EC decay scheme of $^{44}$Ti.}
         \includegraphics[width=\textwidth]{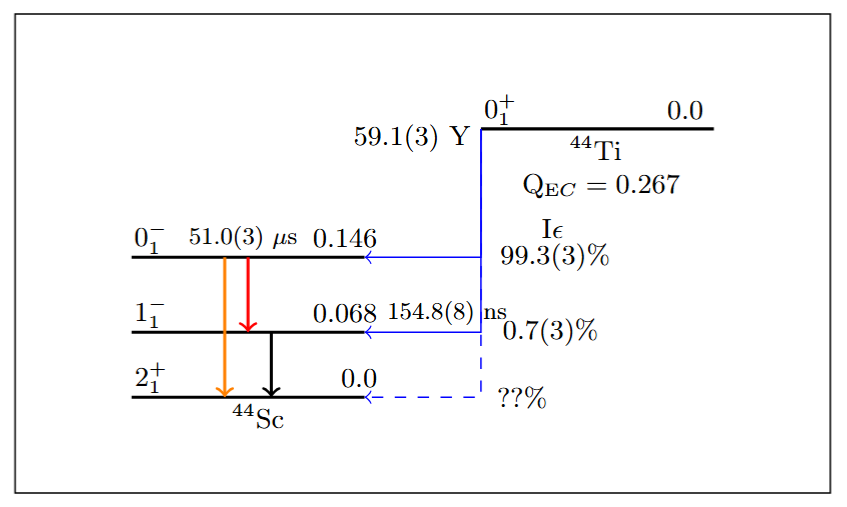}
         \label{fig:44ti}
     \end{subfigure}
     \begin{subfigure}[b]{0.5\textwidth}
         \centering
         \caption{EC decay scheme of $^{57}$Co.}
         \includegraphics[width=\textwidth]{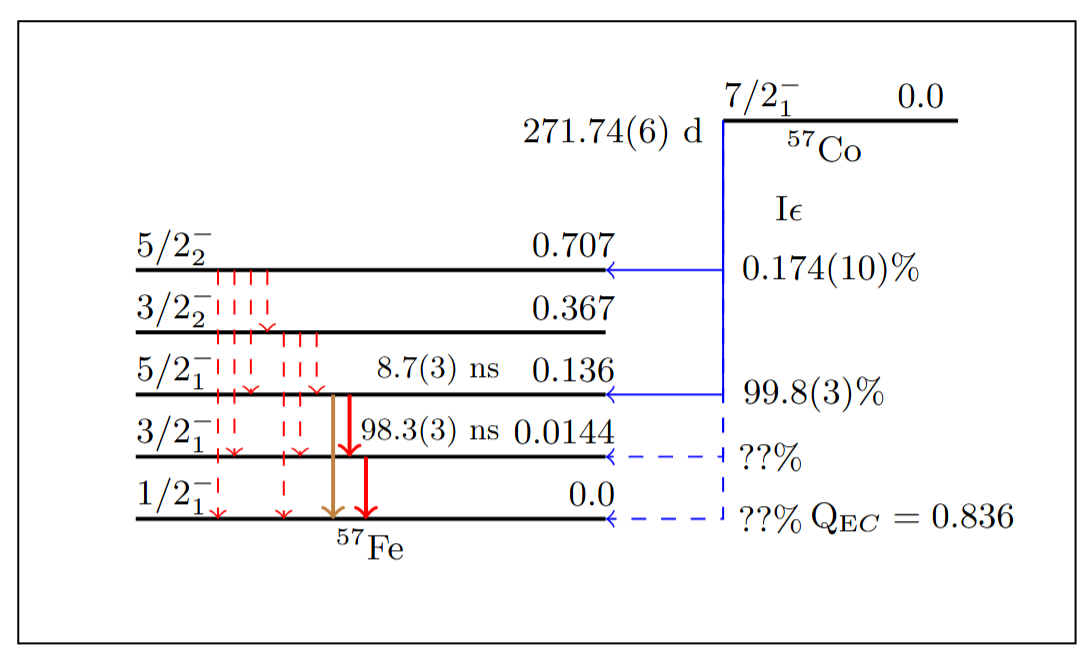}
         \label{fig:57co}
     \end{subfigure}
\end{figure}

Among different proposed detection possibilities, the detection  of axions through magnetic-dipole(M1)-decaying excited states (ESs) of nuclei has been proposed \cite{Donnelly1978}. Here one looks for a missing $\gamma$ in the ALP detector. These experiments have the advantage that they can be performed on the laboratory scale.
The first such detection experiment was proposed by the group of by the group of M. Minowa \cite{axionmain} using $^{139}$Ce radioactive source surrounded by CsI:Tl crystal scintillator to tag X-rays and $\gamma$s. In this experiment, the M1 transition of interest is the $5/2^+_1\rightarrow 7/2^+_1$ 165.86 keV transition shown in Fig. \ref{fig:139ce}. As shown in the figure, the initial $5/2^+_1$ state is fed by an allowed Gamow-Teller (GT) electron-capture (EC) transition from the $3/2^+_1$ ground state of $^{139}$Ce, having a branching ratio (BR) that is practically 1. This EC decay can be detected by its X-ray or Auger-electron (AE) emission, thus tagging the unobservation of an M1 $\gamma$ (or conversion electron (CE) from the internal conversion process) leading to a potential indirect detection of the emission of an ALP. 

A severe experimental problem turns out to be the possible rare branching of the EC decay to the $7/2^+_1$ ground state (GS) of the daughter nucleus $^{139}$La since this transition creates X-rays (or AEs) without emission of $\gamma$-rays, thus mimicking emission of an ALP. This experiment could set the upper limit $\Gamma_{\rm invisible}/\Gamma_{\rm total}< 9.70 \times 10^{-7}$ (95\% C.L.)
to the invisible transition rate relative to the total transition rate. Based on this measurement, an evaluated value of $5(5)\times 10^{-9}$ has been compiled in the NNDC database \cite{NNDC}. The GS transition is very much hindered by the fact that it is a 2$^{nd}$-forbidden non-unique (FNU) EC transition with the emission of an electron neutrino as a higher orbital angular momentum partial wave (see below for more details of these type of transitions). In these laboratory-scale experiments external backgrounds can be significantly lowered by going deep underground and using shielding materials, but the problem of this internal EC background still remains unresolved without dedicated experiments and/or reliable nuclear-structure calculations.

In addition to the case of $^{139}$La, discussed above, there are other two known suitable candidates, namely $^{44}$Sc and $^{57}$Fe, for detection of new dark-matter particles such as ALPs (M1 and M2 transitions) \cite{Donnelly1978}, dark photons (E1 transition) \cite{Filippi2020}, and anapoles (E2 transition) being constituents of anapole dark matter
\cite{Choi2024}. The electromagnetic (EM) decay and EC feeding schemes of these nuclei are depicted Fig. \ref{fig:44ti} and \ref{fig:57co}, respectively. The known EC branches of these sources include the GT transitions $^{57}$Co$(7/2^-_1) \rightarrow \,^{57}$Fe*$(5/2^-_1)$ and $^{57}$Co$(7/2^-_1) \rightarrow \,^{57}$Fe*$(5/2^-_2)$, 
along with the 1$^{st}$-FNU $\beta$ transitions $^{44}$Ti$(0^+_1) \rightarrow \,^{44}$Sc*$(1^-_1)$ and $^{44}$Ti$(0^+_1) \rightarrow \,^{44}$Sc*$(0^-_1)$. In the former case we focus our attention to the EC branch with the strong feeding, namely the allowed transition $^{57}$Co$(7/2^-_1) \rightarrow \,^{57}$Fe*$(5/2^-_1)$, offering interesting M1 and E2 transitions for the studies of exotic dark-matter particles.

There are two major possibilities for dark-matter-particle searches through the EC transition $^{57}$Co$(7/2^-_1) \rightarrow \,^{57}$Fe*$(5/2^-_1)$, where rare unknown EC decays can serve as a background. The first approach involves searches for ALPs through the $^{57}$Fe*$(5/2^-_1) \rightarrow \,^{57}$Fe*$(3/2^-_1)$ M1 transition with a 122.0614 keV $\gamma$ emission, followed by the 
$^{57}$Fe*$(3/2^-_1) \rightarrow \,^{57}$Fe$(1/2^-_1)$ M1 transition with a 14.41300 keV $\gamma$ emission. In the case of an APL signal, the 122.0614 keV $\gamma$ emission would be missing, meaning that only X-rays (or AEs) and 14.41300 keV $\gamma$ emissions would be observed. In this scenario the calculation of the rare ground-state-to-excited-state (GS-to-ES) EC branching $^{57}$Co$(7/2^-_1) \rightarrow \,^{57}$Fe*$(3/2^-_1)$ is crucial, as it will serve as background to the ALP signal. The second approach involves searches for anapoles through the $^{57}$Fe*$(5/2^-_1) \rightarrow \,^{57}$Fe$(1/2^-_1)$ E2 transition with a 136.4743 keV $\gamma$ emission. In the case of an anapole signal, the 136.4743 keV $\gamma$ would be missing, meaning that only 
X-ray (or AE) emissions would be observed. Here, the calculation of the ground-state-to-ground-state (GS-to-GS) EC branch is crucial as it will serve as background for the anapole signal. In addition to the above, the M1-decaying $3/2^-_1$ state is also a potential major source of ALP particles, where the ALP signal can be tagged using the $\gamma$-decays feeding this state.

There are three major possibilities for dark-matter-particle searches through EC transitions from $^{44}$Ti to states in $^{44}$Sc. The first and second possibilities involve the EC transition $^{44}$Ti$(0^+_1) \rightarrow \,^{44}$Sc*$(0^-_1)$. The first approach is searching for ALPs through the $^{44}$Sc*$(0^-_1) \rightarrow \,^{44}$Sc*$(1^-_1)$ M1 transition with a 78.337 keV $\gamma$ emission, followed by the 
$^{44}$Sc*$(1^-_1) \rightarrow \,^{44}$Sc$(2^+_1)$ E1 transition with a 67.875 keV $\gamma$ emission. In the case of an ALP signal, the 78.337 keV $\gamma$ emission would be missing, meaning that only X-rays (or AEs) and 67.875 keV $\gamma$ emissions would be observed. The second approach involves searches of an M2 ALP \cite{Donnelly1978} through the 
$^{44}$Sc*$(0^-_1) \rightarrow \,^{44}$Sc$(2^+_1)$ M2 transition with a 146.212 keV $\gamma$ emission. In the case of the presence of the M2 ALP, the 146.212 keV $\gamma$ emission would be missing, indicating the presence of only X-rays (or AEs). In this case, calculation of the GS-to-GS EC branch is crucial as it will serve as background for the M2-ALP signal.
The third possibility involves searching for a dark photon through the $^{44}$Sc*$(1^-_1) \rightarrow \,^{44}$Sc$(2^+_1)$ E1 transition with a 67.875 keV $\gamma$ emission. In the case of a dark photon signal, the 67.875 keV $\gamma$ emission would be missing, meaning that only X-rays (or AEs) would be observed. As with the second possibility, calculation of the GS-to-GS EC branch is essential as it will serve as background for the dark-photon signal.

Experimentally, there are several critical points to consider for new dark-matter particle searches like those involving ALPs, dark photons and anapoles. First, the detector should have excellent energy resolution to accurately tag low-energy X-rays or AEs. Additionally, precise timing resolution is required to effectively distinguish signals from $\gamma$-rays and X-rays (or AEs) in the transitions of $^{44}$Sc and $^{57}$Fe, as some of these transitions involve nuclear isomeric states, as shown in Figures \ref{fig:44ti} and \ref{fig:57co} respectively. This approach provides
a powerful technique for dramatically reducing radioactive background. Moreover, the veto detector should achieve 100\% efficiency to ensure that 
$\gamma$-rays are not mistakenly identified as new-particle signals. Finally, the detector setup should be located deep underground with low-background lead and copper shielding, similar to WIMP dark-matter searches, to minimize cosmic and environmental background interference.

Our focus in this article is the estimation of BRs for the unknown rare EC branches by using the nuclear shell model (NSM) and a reasonable range $g^{\rm eff}_{\rm A}=0.5-1.27$ of effective values of the weak axial coupling. In addition to the unknown rare EC transitions, we also evaluate the known relevant GS-to-ES EC branches present in the decay of the source isotopes in order to validate our methodology for obtaining reasonable BR estimates.

The non-relativistic limit of the theory of $\beta$ decays is well established and a detailed account is presented in \cite{Behrens1982ElectronRW}. A more concise and practical exposition of the treatment of forbidden $\beta^+$/EC decays is given in 
\cite{ECformalism}. The half-life value for pure EC decays, $t^{\rm EC}_{1/2}$, is given as
$t^{\rm EC}_{1/2}=\kappa/\tilde{C}^{\mathrm{EC}}$, where $\kappa$ takes the value of 6289 s and 
$\tilde{C}^{\mathrm{EC}}$ is the integrated shape function which can be written as \cite{ECformalism}
\begin{equation}
\label{intSF}
    {\tilde{C}}^{\rm EC}=\frac{\pi}{2}\sum_{x=1s,2s}n_x\beta_x^2(p_{\nu_x}/m_ec)^2C(p_{\nu_x})\,,
\end{equation}
when considering the leading channels, namely K capture (1s) and L$_{\mathrm{1}}$ capture (2s). Here $n_x$ are relative occupancies for the respective orbitals, $\beta_x$ are the electron Coulomb amplitudes, and $p_{\nu_x}$ is the momentum of the neutrino when the electron is captured from the atomic orbital $x$ \cite{ECformalism}. The shape factor $C(p_{\nu_x})$ used in Eq.~(\ref{intSF}), containing complicated combinations of phase-space factors and nuclear matrix elements (NMEs), is given for forbidden EC transitions in detail in \cite{ECformalism}. 

Nuclear wave functions for the NMEs and EM observables were computed using the NSM computer program NuShellX$@$MSU \cite{BROWN2014115}. The nuclei $^{44}$Ti and $^{44}$Sc were computed in the SDPFPN model space with the NOWPN interaction and $^{36}$Ar core. The nucleus $^{57}$Co was computed using the FPPN model space and GX1APN interaction. For $^{57}$Fe, KB3GPN interaction was used in the FPPN model space with the truncation scheme of \cite{57fe}. For $^{139}$La and $^{139}$Ce, the JJ55PN model space was used with the SN100PN interaction. Four protons are fixed in the 1g$_{7/2}$ orbital for $^{139}$Ce. 

Comparison of the computed and experimental energies of levels in the nuclei $^{44}$Ti, $^{44}$Sc, $^{57}$Co, $^{57}$Fe, $^{139}$Ce, and $^{139}$La are given in Table I of the Supplemental Material \cite{supplement}. In this material, in Table II, we also compare the computed values of magnetic dipole ($\mu$) and electric quadrupole ($Q$) moments of the involved states with the available data. These moments are indicative of the quality of the wave functions of states involved in the EC and electromagnetic transitions \cite{Suzuki2003,Fujita2018}. In the Supplemental Material we also compare the computed 
M1, E1, M2, and E2 transition probabilities, depicted in Fig.~1, and their mixing ratios, with the available data.

Predicting BRs and (partial) half-lives of $\beta$ transitions using NSM-computed wave functions calls for the use of an effective value $g^{\rm eff}_{\rm A}$ of the weak axial coupling \cite{avcreview}. $g^{\rm eff}_{\rm A}$ is related to the quenching factor $q$ as $q=g^{\rm eff}_{\rm A}/g_A$ \cite{avcreview}, where $g_{\rm A}=1.276$ \cite{gAvalue} is the free-nucleon value. The key inference we consider here is that for different branches of the EC decay of the parent nucleus, $q$ is not unique. This follows from the fact that the $g^{\rm eff}_{\rm A}$ is an artifact reflecting the incompleteness of the wave functions involved \cite{quenchingresolved}. The more complete the state, the closer the value of q is to unity \cite{quenchingresolved}.

The validity of the above inference is justified when we look at the two Gamow-Teller (GT) branches of the decay of $^{57}$Co: The GT transitions that populate the $5/2^-_1$ and $5/2^-_2$ levels in $^{57}$Fe have BRs of 0.998(3) and 0.00174(10), respectively \cite{NNDC}. The branches are identical in a sense that the initial and final spin-parities $J^\pi$ are the same. Also, the wave function of the initial state is common for both. The $g^{\rm eff}_{\rm A}$ for these respective branches are found to be 1.25 and 0.548 for the GT matrix elements $M_{\rm GT}=-0.1065$ and $M_{\rm GT}=0.0572$, respectively, when using the phase-space factors of \cite{ft}. This difference in quenching is due to the difference in the quality of final-state wave functions. In Table~II of the Supplemental Material \cite{supplement}, it can be seen that for the $5/2^-_1$ state, the M1 moment $\mu$ is in very good agreement with the experimental value, justifying a near-unity $q$ value of 0.982. Along the same lines, in the decay of $^{139}$Ce, the GT transition populating the $5/2^+_1$ state with 
BR$\approx 1$ \cite{NNDC} turns out to have $g^{\rm eff}_{\rm A}=0.914$ for $M_{\rm GT}=0.340$. This $g^{\rm eff}_{\rm A}$ value is quite compatible with the relatively well-predicted EM moments of Table~II of the Supplemental Material \cite{supplement}. Based on all these examples, we conjecture that the quenched effective value $g^{\rm eff}_{\rm A}$ appears as a free parameter for all $\beta$ transitions. Therefore, for the aforementioned unknown rare EC branches, we make estimations of BRs and partial half-lives for a conservative interval $g^{\rm eff}_{\rm A}\in [0.5-1.27]$, found sufficient, e.g., in the works \cite{avcreview,113cd,115in}. 

An additional free parameter, namely the small relativistic vector NME (sNME) appears for FNU decays \cite{sNME,113cd,115in}. In the cases where the sNMEs are fitted to experimental BRs, it is seen that there exist two values of sNMEs that reproduce a given BR \cite{sNME1,sNME2,sNME3,115in}. Since three out of four rare EC decays under study are FNU, for these cases values of sNME must be fixed as well. The problem of the sNME is solved by looking at the variation of the EC half-life with the value of sNME for a given $g^{\rm eff}_{\rm A}$. It turns out that for all $g^{\rm eff}_{\rm A}$, there is a unique sNME that maximizes the EC half-life and, in turn, minimizes the corresponding BR. This opens up a way for estimation of the lower limits for the BRs of the rare EC transitions which is valuable for future experiments. 

In order to study closer the influence of the sNME on the BRs we can use as a testing ground the fact that the decay of the 0$^{+}_{1}$ GS of $^{44}$Ti populates the excited states $1^{-}_{1}$ and  $0^{-}_{1}$ in $^{44}$Sc, with branching ratios 0.007(3) and 0.993(3), respectively. Both these branches are 1$^{st}$ FNU but the transition to the $0^{-}_{1}$ state does not depend on sNME \cite{44tiBr} and thus can be left out of the discussion. However, the study of the $1^{-}_{1}$ branch can help locate the sNME that minimizes the BR relative to the two sNMEs that reproduce the experimental BR. The relevant sNME and $g^{\rm eff}_{\rm A}$ values are given in table \ref{44ti sNME}. 
\begin{table}[ht]
\caption{Computed sNME values (sNME$_{1}$ and sNME$_{2}$) which reproduce the experimental BR of the $^{44}$Ti$(0^+_1) \rightarrow ^{44}$Sc*$(1^-_1)$ transition and the sNME value sNME$_{\rm min}$ which minimizes the BR of this transition for the effective weak axial couplings of interest in this work. The CVC value of the sNME, sNME$_{\rm CVC}$, is given for reference.}
    \label{44ti sNME}
\begin{ruledtabular}
    \begin{tabular}{cccc}
        $g^{\rm eff}_{\rm A}$ & sNME$_1$ & sNME$_2$ & sNME$_{\rm min}$ \\ \midrule
        0.5 & -0.00535 & -0.00094 & -0.00314 \\
        0.6  & -0.00596 & -0.00154 & -0.00375 \\
        0.7 & -0.00656 & -0.00215 & -0.00435 \\
        0.8 & -0.00717 & -0.00275 & -0.00496 \\
        0.9 & -0.00777 & -0.00336 & -0.00556 \\
        1.0 & -0.00838 & -0.00396 & -0.00617 \\
        1.1 & -0.00898 & -0.00457 & -0.00678 \\
        1.2 & -0.00959 & -0.00517 & -0.00738 \\
        1.27 & -0.01001 & -0.00560 & -0.00780 \\
         & sNME$_{\rm CVC}=-0.01350$ & \\
            \end{tabular}
\end{ruledtabular}  
\end{table} 

\begin{table*}[ht]
\caption{\label{BRandhf} Values of sNMEs (sNME$_{\rm min}$) which minimize the rare BR (BR$_{\rm min}$) (and hence maximize the corresponding half-life $t_{\rm 1/2,max}$) for the effective weak axial couplings of interest in this work. The transition $^{57}$Co$(7/2^-_1)\rightarrow\,^{57}$Fe$(1/2^-_1)$ is FU and thus independent of sNME. The lowest values of BR are highlighted, and they can serve as conservative estimates for experiments. 
}
    \begin{ruledtabular}
    \begin{tabular}{c|cc|cc|c|cc}   
    & \multicolumn{2}{c}{$^{44}$Ti$(0^+_1)\rightarrow\,^{44}$Sc$(2^+_1)$} & \multicolumn{2}{c}{$^{57}$Co$(7/2^-_1)\rightarrow\,^{57}$Fe*$(3/2^-_1)$} & $^{57}$Co$(7/2^-_1)\rightarrow\,^{57}$Fe$(1/2^-_1)$ & \multicolumn{2}{c}{$^{139}$Ce$(3/2^+_1)\rightarrow\,^{139}$La$(7/2^+_1)$}\\
    & \multicolumn{2}{c}{2$^{nd}$-FNU}& \multicolumn{2}{c}{2$^{nd}$-FNU}& 2$^{nd}$-FU & \multicolumn{2}{c}{2$^{nd}$-FNU} \\ \midrule
    $g^{\rm eff}_{\rm A}$ & sNME$_{\rm min}$ & BR$_{\rm min}$ & sNME$_{\rm min}$ & BR$_{\rm min}$ & BR&  sNME$_{\rm min}$ & BR$_{\rm min}$ \\ \midrule
    0.5 & -0.0203 & 6.15E-9 & 0.0293 & 2.52E-8 & \textbf{1.63E-9} & -0.136 & \textbf{1.00E-9}  \\
    0.6 & -0.0238 & 5.95E-9 & 0.0343 & \textbf{2.47E-8} &  2.35E-9 & -0.163 & 1.12E-9  \\
    0.7 & -0.0272 & 5.76E-9 & 0.0393 & \textbf{2.47E-8} & 3.19E-9 & -0.191 & 1.25E-9  \\
    0.8 & -0.0306 & 5.57E-9 & 0.0442 & 2.52E-8 & 4.17E-9 & -0.219 & 1.38E-9  \\
    0.9 & -0.0341 & 5.38E-9 & 0.0492 & 2.61E-8 & 5.28E-9 & -0.246 & 1.53E-9  \\
    1.0 & -0.0375 & 5.20E-9 & 0.0542 & 2.74E-8 & 6.52E-9 & -0.274 & 1.68E-9  \\
    1.1 & -0.0409 & 5.02E-9 & 0.0592 & 2.91E-8 & 7.89E-9 & -0.302 & 1.84E-9  \\
    1.2 & -0.0444 & 4.84E-9 & 0.0642 & 3.13E-8 & 9.39E-9 & -0.330 & 2.01E-9  \\
    1.27 & -0.0468 & \textbf{4.72E-9} & 0.0676 & 3.31E-8 & 1.05E-8 & -0.349 & 2.13E-9 \\
    \end{tabular}
    \end{ruledtabular} 
\end{table*}

In table \ref{44ti sNME} it can be seen that sNME$_{\rm min}$ is always located in between NME$_{1}$ and NME$_{2}$. This means that sNME values larger or smaller than sNME$_{\rm min}$ produce larger branchings for all relevant $g^{\rm eff}_{\rm A}$. The CVC value of sNME, denoted as sNME$_{\rm CVC}$ in Table \ref{44ti sNME}, is also unique and can potentially give additional information on the physical value of sNME \cite{sNME,113cd,115in}. Yet, due to the lack of perfectly modeled wave functions, predictions using the CVC value usually fall short of reproducing experimental values of BRs, making the CVC value only a good reference in searches of the proper value of the sNME in different nuclear models \cite{sNME,113cd,115in}. This is also the case for the 1$^{st}$ FNU transition at hand, where 
sNME$_{\rm CVC}$ reproduces the experimental BR for $g^{\rm eff}_{\rm A}>1.276$.

For the rare EC transitions relevant for this work, BRs are tabulated in Table~\ref{BRandhf}. There we give the value sNME$_{\rm min}$, minimizing the rare BR (BR$_{\rm min}$) (and maximizing the half-life $t_{\rm 1/2,max}$), for the FNU GS-to-GS decays of $^{44}$Ti and $^{139}$Ce, and the FNU GS-to-ES and FU GS-to-GS decays of $^{57}$Co. As can be seen, the BR$_{\rm min}$ changes smoothly with $g^{\rm eff}_{\rm A}$ being smallest for $g^{\rm eff}_{\rm A}=0.5$ for the GS-to-GS decays of $^{57}$Co and $^{139}$Ce, whereas for the GS-to-GS decay of $^{44}$Ti the smallest value is achieved for $g^{\rm eff}_{\rm A}=1.27$. For the GS-to-ES decay of $^{57}$Co BR$_{\rm min}$ is achieved for a $g_{\rm A}$ value $g^{\rm eff}_{\rm A}=0.65$ (in Table~\ref{BRandhf} the rounded-up numbers become the same for $g^{\rm eff}_{\rm A}=0.6$ and $g^{\rm eff}_{\rm A}=0.7$). Therefore, the most conservative estimates are the lowest BRs obtained, $4.72\times 10^{-9}$ (with the corresponding maximum half-life 
$t_{\rm 1/2,max}=3.95\times 10^{17}$ s) for the GS-to-GS decay of $^{44}$Ti$(0^+_1)$, $1.63\times 10^{-9}$ ($t_{\rm 1/2,max}=1.44\times 10^{16}$ s) for the GS-to-GS decay of $^{57}$Co$(7/2^-_1)$, $2.47\times 10^{-8}$ ($t_{\rm 1/2,max}=9.51\times 10^{14}$ s) for the GS-to-ES decay of $^{57}$Co$(7/2^-_1)$, and $1.00\times 10^{-9}$ ($t_{\rm 1/2,max}=1.19\times 10^{16}$ s) for the GS-to-GS decay of $^{136}$Ce$(3/2^+_1)$,
and these are the ones suggested for the experiments. It should be noted here that for $^{139}$Ce our result BR$_{\rm min}=1.00\times 10^{-9}$ is well in line with the measured upper limit $\Gamma_{\rm invisible}/\Gamma_{\rm total}< 9.70 \times 10^{-7}$ (95\% 
C.L.) \cite{axionmain}, and compiled value of $5(5)\times 10^{-9}$ \cite{NNDC}. 

In summary, we give for the first time theoretical estimates for the minimum branching ratios of the rare ground-state-to-ground-state second-forbidden electron-capture transitions
$^{44}$Ti$(0^+_1)\rightarrow\,^{44}$Sc$(2^+_1)$, 
$^{57}$Co$(7/2^-_1)\rightarrow\,^{57}$Fe$(1/2^-_1)$, and $^{139}$Ce$(3/2^+_1)\rightarrow\,^{139}$La$(7/2^+_1)$, and for the rare ground-state-to-excited-state second-forbidden electron-capture transition 
$^{57}$Co$(7/2^-_1)\rightarrow\,^{57}$Fe*$(3/2^-_1)$. The nuclear-structure calculations have been done exploiting well-established nuclear shell-model Hamiltonians and an advanced theory of $\beta$ decay. The obtained minimum branching ratios are $4.72\times 10^{-9}$ for the decay of Ti, $1.63\times 10^{-9}$ (GS-to-GS) and $2.47\times 10^{-8}$ (GS-to-ES) for the decays of Co, and $1.00\times 10^{-9}$ for the decay of Ce, the last fully compatible with the available experimental upper limit of the branching ratio. It is conjectured that in the absence of experimental measurements of these decay branches, these estimates are of utmost importance for terrestrial searches of exotic new particles, like axionic dark matter, dark photons, and anapole dark matter exploiting nuclear transitions, as they can potentially mimic detection rates of these particles in the related experiments.

\begin{table}[ht]
    \caption{Glossary}
    \centering
    \begin{ruledtabular}
    \begin{tabular}{cc}
        Acronym/symbol & Defination \\ \midrule
        AE & Auger electron \\
        ALP &  Axion-like particle  \\
        BR     &  Branching ratio  \\
        BSMP & Beyond the standard model physics \\
        CE & Conversion electron \\
        CVC & Conserved vector current \\
        EC     & Electron capture \\
        EM & Electromagnetic \\
        ES     & Excited state   \\
        FNU &  Forbidden non-unique  \\
        FU  &  Forbidden unique  \\
        GS     & Ground state   \\
        GT     &  Gamow-Teller  \\
        NSM  &  Nuclear shell model  \\
        M1     &  Magnetic Dipole  \\
        $\mu$  &  Magnetic dipole moment \\
        NME   &  Nuclear matrix element  \\
        $Q$  & Electric quadrupole moment \\
        sNME  & small relativistic vector NME \\ 
        sNME$_{\rm min}$ &  sNME that minimizes the BR \\
        sNME$_{\rm CVC}$ &  CVC value of sNME \\ 
    \end{tabular}              
    \end{ruledtabular}
    \label{Glossary}
\end{table}
\begin{acknowledgments}
J.S. acknowledges support from project PNRR-I8/C9-CF264, Contract No. 760100/23.0
5.2023 of the Romanian Ministry of Research, Innovation and Digitization (the NEPTUN project). H.J. Kim acknowledges support by the National Research Foundation of Korea (NRF) grant, funded by the Korean government (MSIT), Contract No. RS-2024-00348317.
We acknowledge grants of computer capacity from the Finnish Grid and
Cloud Infrastructure (persistent identifier urn:nbn:fi:research-infras-2016072533 ) and the support by CSC – IT Center for Science, Finland, for the generous computational resources.
\end{acknowledgments}


\bibliography{axions}

\end{document}


\title{Supplemental Material for: Constraints for rare electron-capture decays mimicking detection of dark-matter particles in nuclear transitions}

\maketitle

Here we give information related to the quality of the computed wave functions, namely we compare the values of the computed energies and electromagnetic observables of nuclear states involved in the EC and electromagnetic transitions depicted in Fig.~1 of the main text \cite{PRL-Agni}.

\begin{table}[h]
\caption{\label{Levels and rates}
Experimental and computed values of energies for the states involved in EC decays in the three axion sources. Energies are given in MeV. Experimental values are taken from \cite{NNDC}.}
\begin{ruledtabular}
\begin{tabular}{cccc}
 Nucleus & $J^{\pi}$ & E$_{\rm exp.}$ & E$_{\rm theo.}$ \\ \midrule
$^{44}_{22}$Ti$^{}_{22}$ & $0^{+}_{1}$ & 0 & 0  \\ 
$^{44}_{21}$Sc$^{}_{23}$ & $2^{+}_{1}$ & 0 & 0  \\
  & $1^{-}_{1}$ & 0.068 & 7.410  \\
  & $0^{-}_{1}$ & 0.146 & 9.093  \\
$^{57}_{27}$Co$^{}_{30}$ & $7/2^{-}_{1}$ & 0 & 0 \\ 
$^{57}_{26}$Fe$^{}_{31}$ & $1/2^{-}_{1}$ & 0 & 0  \\
  & $3/2^{-}_{1}$ & 0.0144 & 0.022  \\
  & $5/2^{-}_{1}$ & 0.136 & 0.063 \\ 
  & $5/2^{-}_{2}$ & 0.706 & 0.820 \\ 
$^{139}_{58}$Ce$^{}_{81}$ & $3/2^{+}_{1}$ & 0 & 0  \\ 
$^{139}_{57}$La$^{}_{82}$ & $7/2^{+}_{1}$ & 0 & 0  \\
  & $5/2^{+}_{1}$ & 0.166 & 0.082   \\ 
\end{tabular}
\end{ruledtabular}
\end{table}

\begin{table}[]
\caption{\label{Electromagnetic moments}
Experimental and theoretical values of EM moments of the states involved in the EC decays to the three axion sources. The magnetic dipole ($\mu$) and electric quadrupole ($Q$) moments are given in units of nuclear magnetons ($\mu_N/c$) and $e$barns, respectively. In the calculations, an effective charge 1.5$e$ (0.5$e$) for proton (neutron) and bare $g$ factors were used \cite{Suhonen2007}. Experimental values are taken from \cite{NNDC}.  }
\begin{ruledtabular}
\begin{tabular}{cccccc}
 Nucleus & $J^{\pi}$ & $\mu_{\rm exp}$ & $\mu_{\rm theo}$ & $Q_{\rm exp}$ & $Q_{\rm theo}$ \\ \midrule
$^{44}_{22}$Ti$^{}_{22}$ & $0^{+}_{1}$ & 0 & 0 & 0 & 0 \\ 
$^{44}_{21}$Sc$^{}_{23}$ & $2^{+}_{1}$ & +2.498(5) & +2.605 & +0.10(5) & +0.0806 \\ 
  & $1^{-}_{1}$ & +0.342(6) & -1.287 & +0.21(2) & +0.0750 \\
  & $0^{-}_{1}$ & 0 & 0 & 0 & 0 \\
$^{57}_{27}$Co$^{}_{30}$ & $7/2^{-}_{1}$ & +4.720(10) & +4.568 & +0.54(10) & +0.3876 \\ 
$^{57}_{26}$Fe$^{}_{31}$ & $1/2^{-}_{1}$ & +0.09064(7) & +0.130 & 0 & 0 \\
  & $3/2^{-}_{1}$ & -0.15531(18) & -0.387 & +0.160(8) & +0.1726 \\
  & $5/2^{-}_{1}$ & +0.935(10) & +0.980 & - & -0.3173\\ 
  & $5/2^{-}_{2}$ & - & +0.533 & - & -0.0533  \\ 
$^{139}_{58}$Ce$^{}_{81}$ & $3/2^{+}_{1}$ & +1.06(4) & +1.316 & - & +0.2523 \\
$^{139}_{57}$La$^{}_{82}$ & $7/2^{+}_{1}$ & +2.7791(2) &  +1.756  & +0.206(4) & +0.1267 \\
  & $5/2^{+}_{1}$ & - & +4.668 & - & -0.1698 \\ 
\end{tabular}
\end{ruledtabular}
\end{table}

\begin{table}[]
\caption{\label{M1rates} Reduced transition probabilities in Weisskopf units (W.u.) for the electric/magnetic dipole (E1/M1) and electric/magnetic quadrupole (E2/M2) transitions, and their E2/M1 and M2/E1 mixing ratios $\delta$, for the transitions accompanying the exotic/dark-matter particle de-excitation under consideration. The used effective charges and $g$ factors are those listed in Table~\ref{Electromagnetic moments}. The infinity "$\infty$" comes from division by zero (transition forbidden by angular-momentum considerations). Experimental values are taken from \cite{NNDC}.}    
    \centering
    \begin{ruledtabular}
        \begin{tabular}{cccc}
         Transition & $B$(M1) & $B$(E2) & $\delta$ \\ \midrule
         $^{44}$Sc($0^{-}_{1}\rightarrow 1^{-}_{1}$) & & &\\
         theo & 0.01752 & 0 & 0 \\
         exp & $8.71(5)\times 10^{-7}$  & - & -  \\
         $^{57}$Fe($3/2^{-}_{1}\rightarrow 1/2^{-}_{1}$) & & &\\
         theo & 0.03545 & 0.49 & 0.00183 \\
         exp & 0.0078(3) & 0.37(7) & 0.00223(18)  \\   
         $^{57}$Fe($5/2^{-}_{1}\rightarrow 3/2^{-}_{1}$) & & &\\
         theo & $9.27\times 10^{-4}$ & 0.57 & 0.02 \\
         exp & 0.00118(17) & 2.3(4) & 0.120(1)  \\
         $^{57}$Fe($5/2^{-}_{1}\rightarrow 1/2^{-}_{1}$) & & &\\
         theo & 0 & 13.323 & $\infty$ \\
         exp & - & 10.9(16) & - \\
         $^{139}$La($5/2^{+}_{1}\rightarrow 7/2^{+}_{1}$) & & &\\
         theo & 0.0001795 & 0.04414 & 0.052 \\ 
         exp & 0.00257(4) & - & -  \\ \midrule
        
         Transition & $B$(E1) & $B$(M2) & $\delta$ \\ \midrule 
          
         $^{44}$Sc($1^{-}_{1}\rightarrow 2^{+}_{1}$) & & &\\
         theo & 0.01232 & 0.432 & 0.02 \\
         exp & $1.034(6)\times 10^{-5}$  & - & -  \\
         $^{44}$Sc($0^{-}_{1}\rightarrow 2^{+}_{1}$) & & &\\
         theo & 0 & 0.0221 & $\infty$ \\
         exp & - & 0.000673(26) & -  \\

    \end{tabular}
    \end{ruledtabular}
    
\end{table}

Experimental and computed energies of states involved in all the discussed EC and electromagnetic decays are given in Table~\ref{Levels and rates}. The values of magnetic dipole ($\mu$) and electric quadrupole ($Q$) moments of the involved states are given in Table~\ref{Electromagnetic moments}. In addition to this, transition strengths of M1, E1, M2, and E2 transitions, and their mixing ratios, are compared with data in Table~\ref{M1rates}.

We compare the computed and experimental energies of levels involved in the EC and electromagnetic tramsitions in Table~\ref{Levels and rates}. One can see that the ground states are correctly predicted by the NSM. The energies of the excited states are reasonably reproduced, except for the $1^-_1$ and $0^-_1$ states in $^{44}$Sc for which the computed energies are way too high. This discrepancy in the excitation energies is clearly reflected also as badly described EM moments of these states, as seen in Table~\ref{Electromagnetic moments}. On the other hand, the computed EM moments are quite compatible with the experimental ones for the other considered states, in particular for those involved in the unknown EC transitions of interest for this work. Since the key spectroscopic observables for the states relevant for the EC studies are satisfactory, we believe that the computed estimates for the minimum EC BRs for $^{44}$Ti in Table \ref{Electromagnetic moments} of the main text are of correct order of magnitude.

Considering the reduced transition probabilities displayed in Table~\ref{M1rates} one can state that large deviations between the computed and measured reduced transition probabilities are observed for the transitions in $^{44}$Sc. This is probably due to the inaccurate wave functions of the 
$1^-_1$ and $0^-_1$ states, as indicated by their unrealistically large energies in Table~\ref{Levels and rates}.

\bibliography{Axions-Supplemental_Material}